\documentclass[amssymb,aps,prl,twocolumn,groupedaddress,showpacs]{revtex4}

\usepackage{graphicx}

\begin{document}

\title{Pulling absorbing and collapsing polymers from a surface}
\author{J. Krawczyk}
\email{krawczyk.jaroslaw@tu-clausthal.de}
\affiliation{Institut f\"ur Theoretische Physik, Technische Universit\"at Clausthal, Arnold Sommerfeld Stra\ss e 6, D-38678 Clausthal-Zellerfeld, Germany}
\author{A. L. Owczarek}
\email{aleks@ms.unimelb.edu.au}
\affiliation{Department of Mathematics and Statistics, The University of Melbourne, 3010, Australia}
\author{T. Prellberg}
\email{thomas.prellberg@tu-clausthal.de}
\affiliation{Institut f\"ur Theoretische Physik, Technische Universit\"at Clausthal, Arnold Sommerfeld Stra\ss e 6, D-38678 Clausthal-Zellerfeld, Germany}
\author{A. Rechnitzer}
\email{andrewr@ms.unimelb.edu.au}
\affiliation{Department of Mathematics and Statistics, The University of Melbourne, 3010, Australia}

\date{\today }

\begin{abstract}

A self-interacting polymer with one end attached to a sticky surface
has been studied by means of a flat-histogram stochastic growth
algorithm known as FlatPERM.  We examined the four-dimensional
parameter space of the number of monomers up to 91, self-attraction,
surface attraction and force applied to an end of the polymer. Using
this powerful algorithm the \emph{complete} parameter space of
interactions and force has been considered. Recently it has been
conjectured that a hierarchy of states appears at low temperature/poor
solvent conditions where a polymer exists in a finite number of layers
close to a surface. We find re-entrant behaviour from a stretched phase
into these layering phases when an appropriate force is applied to the
polymer.  We also find that, contrary to what may be expected, the
polymer desorbs from the surface when a sufficiently strong critical
force is applied and does \emph{not} transcend through either a series
of de-layering transitions or monomer-by-monomer transitions.

\end{abstract}

\maketitle

New experimental methods in the physics of macromolecules
\cite{strick2001a-a} have been used to study and manipulate single
molecules and their interactions.  These methods make a contribution
to our understanding of such phenomena as protein folding or DNA
un-zipping; one can push or pull a single molecule and watch how it
responds.  It is possible to apply (and measure) forces large enough
to induce structural deformation of single molecules.  One can monitor
the mechanism of some force-driven phase transition occurring at the
level of a single molecule. Theoretical understanding of this
behaviour has attracted much attention
\cite{marenduzzo2003a-a,rosa2003a-a,orlandini2004a-a}.  

The response of a single polymer to an external force under good
solvent conditions was considered some time
ago\cite{gennes1979a-a}. The response under poor solvent conditions
(below the $\theta$-point), where the self-attraction and an external
force compete with each other, was examined later
\cite{halperin1991a-a,lai1996a-a,grassberger2002a-a,marenduzzo2003a-a,rosa2003a-a}.
Another phenomenon commonly studied in polymer physics is the
adsorption of a polymer tethered to a ``sticky'' wall. The response of
such a polymer to a force perpendicular to the wall has also recently
been considered
\cite{orlandini2004a-a,krawczyk2004=a-:a,mishra2004a-a}.
However, when both the self-attraction (ie.\ monomer-monomer
attraction that leads to polymer collapse) and the surface attraction
(ie monomer-wall attraction that leads to adsorption) compete the
response to an external force has not yet been elucidated (some
interesting results can be found in
\cite{celestini2004=a-a}). Certainly, the full phase
diagram has not been considered. Making such a study now is all the
more timely because of the very recent discovery
\cite{krawczyk2004=b-:a} of a new low temperature phenomenon of
layering transitions (without a force). It is this layering phenomenon
that raises the intriguing question about the response a
low-temperature polymer may have to an external force. In the layering
state a polymer is tightly confined within a fixed number of layers
above the wall. It may be therefore be especially interesting
experimentally to examine such a situation.

We demonstrate for the \emph{first} time how the full two-dimensional
phase diagram of surface and self-attraction changes as the force is
increased. The desorbed extended regime, which changes its scaling
behaviour as soon as the force is made non-zero, simply grows as the
force is increased.  The second-order phase transitions of adsorption
and collapse become first order. The rest of the phase diagram remains
relatively unaffected as long as the force is small. After the force
passes a critical value, which depends on the zero temperature force
required to pull a polymer from a wall, a re-entrant behaviour occurs
at low temperatures. For different values of the force, this
re-entrant behaviour occurs for both the adsorption and collapse of
polymers, including the layering phases mentioned above. We provide a
full force-temperature diagram for all ratios of surface attraction to
self-attraction. All this is achieved with the use of a recently
developed algorithm, FlatPERM \cite{prellberg2004a-a}, that is
specifically designed to obtain information about the whole phase
diagram in one simulation run: it is effectively a stochastic
enumeration algorithm that estimates the complete density of states.

The model we have considered is a self-avoiding walk on a
three-dimensional cubic lattice in a half-space. The
self-avoiding walk is attached at one end to the wall with surface
energy per monomer of $\varepsilon_s$ for
\emph{visits} to the wall. The self-avoiding walk
self-interacts via a nearest-neighbour energy of attraction
$\varepsilon_b$ per monomer-monomer
\emph{contact}.    (Note that the attractive energies 
$\varepsilon_b$ and $\varepsilon_s$ are taken to be positive.) A force
$f$ is applied in the direction perpendicular to the boundary of the
half-space (wall).  The total energy of a configuration $\varphi_n$ of
length (number of monomers) $n$ is given by
\begin{equation}
E_n(\varphi_n) = -m_b(\varphi_n) \varepsilon_b - m_s(\varphi_n)
\varepsilon_s - f h(\varphi_n)
\end{equation}
and depends on the number of non-consecutive nearest-neighbour pairs
(contacts) along the walk $m_b$, the number of visits to the planar
surface $m_s$, and the height $h$ in the direction perpendicular to
the boundary (wall) of the half-space.  Figure 1 shows a diagram of
the two-dimensional analogue.
\begin{figure}[ht!]
\begin{center}
\includegraphics[width=8cm]{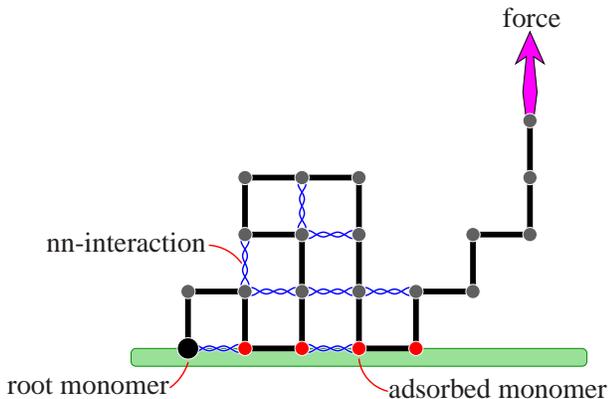}
\vspace{1mm}
\caption{\it A diagram showing the two-dimensional version of the
three-dimensional model simulated.}
\end{center}
\end{figure}
For convenience, we define $\beta_b= \varepsilon_b/k_BT$, $\beta_s=
\varepsilon_s/k_BT$ and $\beta_f= f/k_BT $ where $T$  is  the temperature and
$k_B$ is the Boltzmann constant.  The partition function is given by
\begin{equation}
Z_n(\beta_b,\beta_s,\beta_f)=
\sum_{m_b,m_s,h} C_{n,m_b,m_s,h} e^{\beta_b m_b+\beta_s m_s + \beta_f
h}
\end{equation}
 with
$C_{n,m_b,m_s,h}$ being the density of states. It is this density of
states that is estimated directly by the FlatPERM simulation. Our
algorithm grows a walk monomer-by-monomer starting on the surface. We
obtained data for each value of $n$ up to $n_{max}=91$, and all permissible
values of $m_b$, $m_s$, and $h$.

When $f=0$ the phase diagram of the model contains several phases and
transitions between them
\cite{veal1991a-a,vrbova1996a-a,rajesh2002a-a,krawczyk2004=b-:a}.
\begin{figure}[ht!]
\begin{center}
\begin{tabular}{cc}
(a) & \\
(b) & \includegraphics[width=78mm]{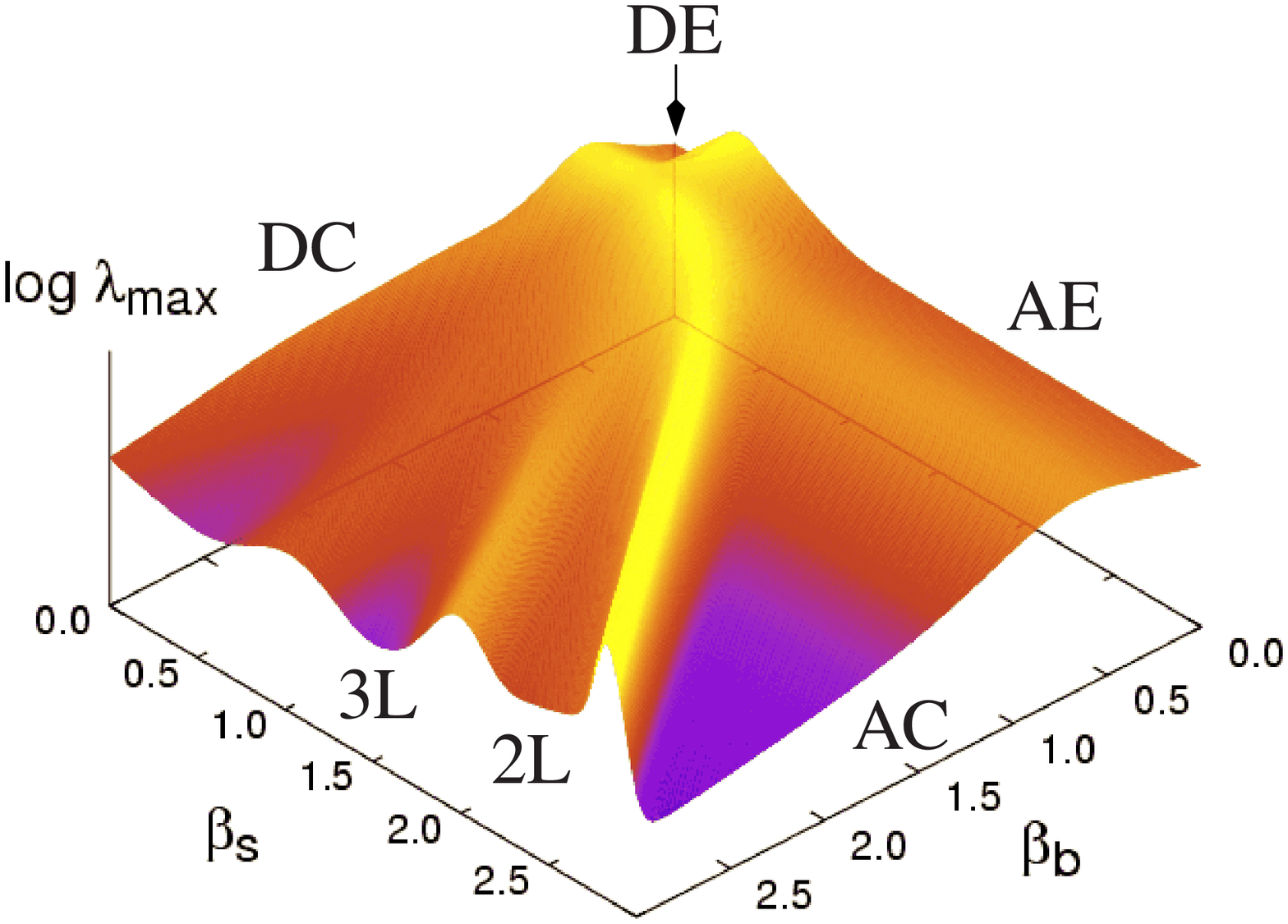}\\
(c) & \includegraphics[width=78mm]{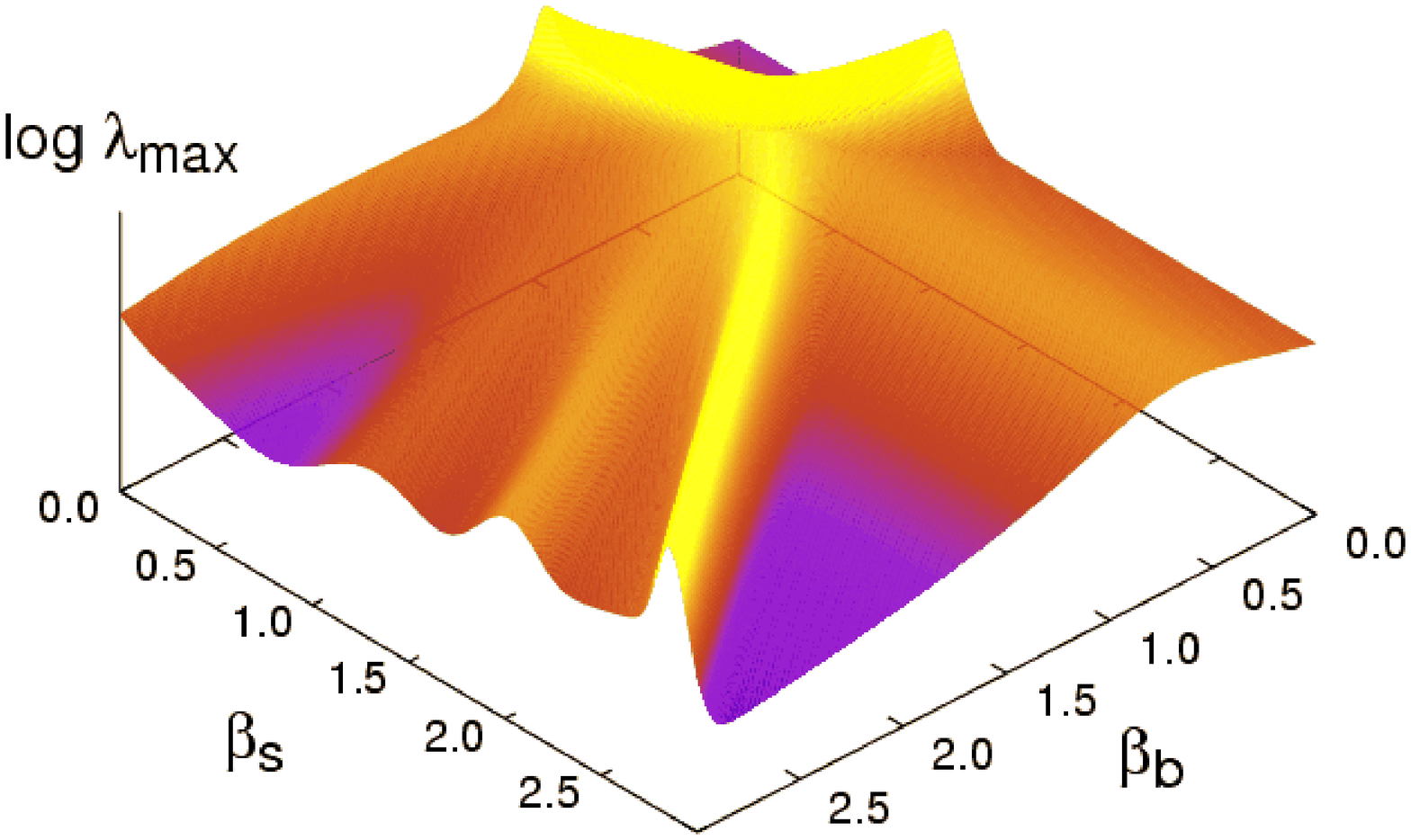}\\
& \includegraphics[width=78mm]{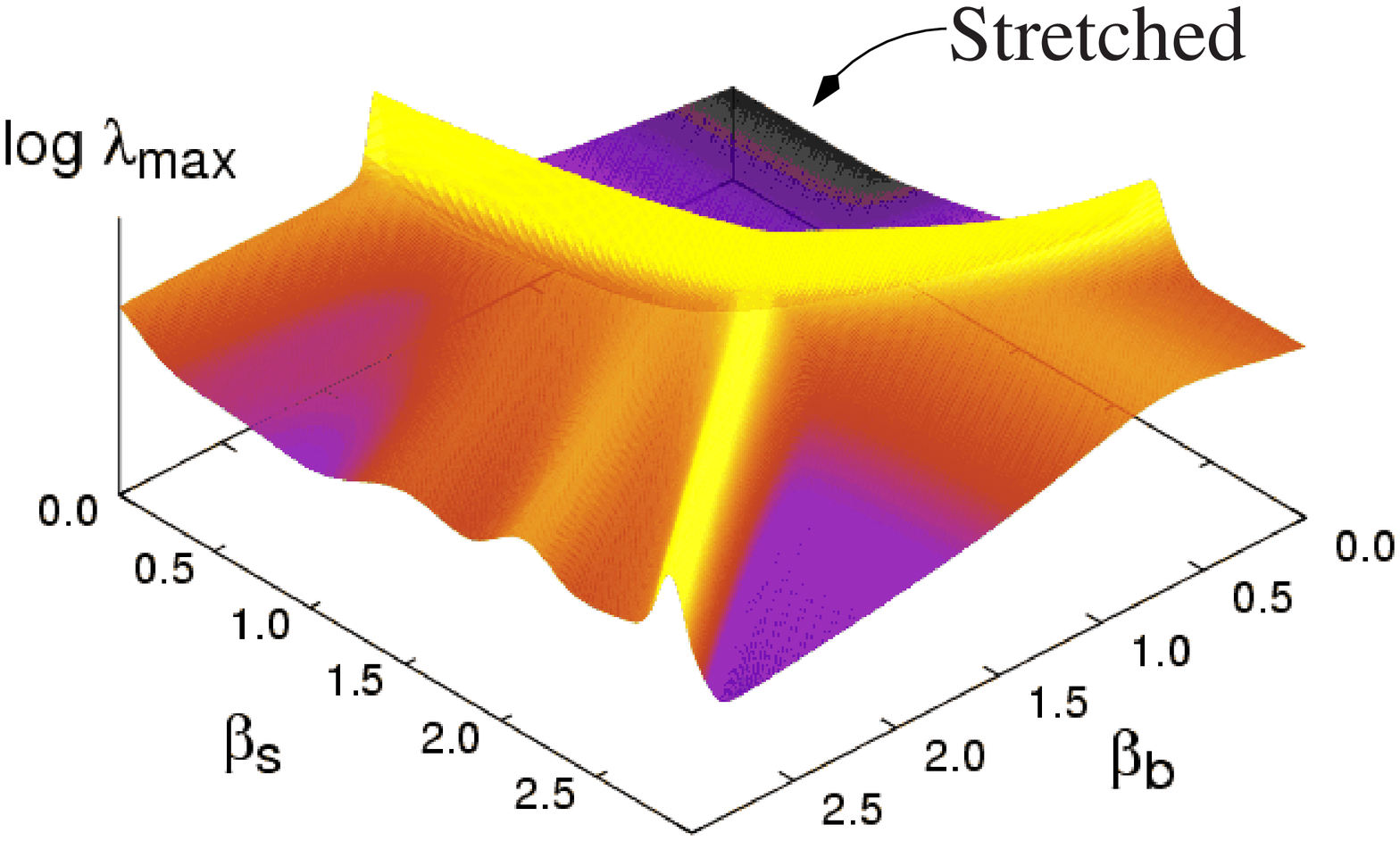}
\end{tabular}
\vspace{1mm}
\caption{\it Plots of the logarithm of the maximum eigenvalue of matrix of second
derivatives of the free energy for three values of $\beta_f=0.0, 1.5
\mbox{ and } 3.0$. In the top far corner of the plot is the location
of the desorbed-extended phase when $\beta_f=0$ and the `Stretched'
phase when $\beta_f>0$. The location of the 2-layer (2L), 3-layer
(3L), adsorbed-collapsed (AC) and adsorbed-extended AE phases do not
seems to move greatly as $\beta_f$ is increased.}
\end{center}
\end{figure}
For small $\beta_b$ and $\beta_s$ there is a desorbed extended (DE)
phase with the polymer behaving as a free flexible polymer in solution
(ie.\ swollen or extended in three dimensions). For $\beta_b$ fixed
and small, increasing $\beta_s$ leads to a second-order phase
transition (adsorption) to a state in which the polymer is adsorbed
onto the wall and behaves in a swollen two-dimensional fashion
(AE). Alternately, if $\beta_b$ is increased at small $\beta_s$ a
second-order collapse transition occurs to a state resembling a dense
liquid drop. This phase is known as desorbed collapsed (DC) on the
assumption that it has little contact with the wall
\cite{veal1991a-a,vrbova1996a-a}. However, it has been subsequently
argued \cite{singh2001a-a} that for larger $\beta_b$ there is also a
polymer-surface transition to a Surface-Attached Globule (SAG) phase,
where the polymer behave as a liquid drop partially wetting the
wall. This transition will not be seen directly by studying
thermodynamic polymer quantities as it occurs as a singularity in the
surface free energy and not the bulk free energy of the polymer. When
$\beta_s$ is large, so that the polymer is adsorbed onto the wall,
increasing $\beta_b$ will result in a two-dimensional (second-order)
transition to a adsorbed and collapsed phase (AC). In recent work
\cite{krawczyk2004=b-:a} this AC phase was also referred to as the
\emph{1-layer} phase because for very large $\beta_b$ and $\beta_s <
\beta_b$ there exist meta-stable $\ell$-layer phases where the polymer
is two-dimensionally collapsed and more-or-less restricted to $\ell$
layers parallel to the wall (for small $\ell$). A series of
first-order transitions between adjacent values of $\ell$ occur as
$\beta_s$ is varied at fixed $\beta_b$. All these transition lines can
be seen in the Figure 2 (a) which shows a plot of the logarithm of the
maximum eigenvalue of the ($2\times 2$) matrix of second derivatives
in the variables $\beta_b$ and $\beta_s$ of
$\log(Z_n(\beta_b,\beta_s,\beta_f))$ for fixed $\beta_f=0$. The local
maxima indicate transitions. The transition to the AC phase is
expected to be first order in the thermodynamic limit.

Using the evidence available in the literature 
\cite{grassberger2002a-a,marenduzzo2003a-a,rosa2003a-a,orlandini2004a-a,krawczyk2004=a-:a,mishra2004a-a}, let us now
consider what we can expect when $f>0$. The first important feature to
note is that the isotropic DE phase is replaced by an anisotropic
phase in which the height of the end point of the polymer scales
linearly with $n$; we denote this phase as the \emph{Stretched} phase.
Consequently the transition from stretched to adsorbed phases becomes
first-order \cite{orlandini2004a-a}. Likewise, at least in three
dimensions \cite{grassberger2002a-a}, the transition from the
vertically stretched phase to the collapsed phase also becomes
first-order. This implies that the multicritical point (where for
$f=0$ the DE, AE and DC phases meet) is now a triple point: the
meeting of three first order lines. 
The transition from the AE to AC
phases should not be effected by the application of a small force as
it acts in a direction perpendicular to the driving phenomenon of
planar collapse.
\begin{figure}[bh!]
\begin{center}
\includegraphics[width=8cm]{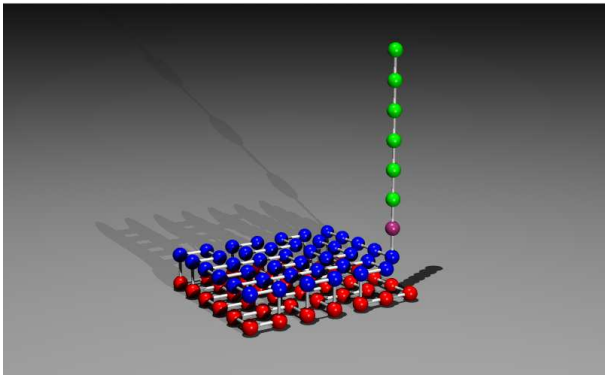}
\vspace{1mm}
\caption{\it A typical configuration resulting from the application of 
the critical force $f_c$ to a polymer in the 2-layer adsorbed
collapsed phase.}
\end{center}
\end{figure}
 Finally, it is intriguing to ask what happens to the
layering phases observed in
\cite{krawczyk2004=b-:a}.
One can imagine that the force simply extends a vertical `tail' from a
layered block (see figure 3) and that as the force is increased the
monomers are peeled off one at a time with corresponding
micro-transitions \cite{celestini2004=a-a} for each monomer pulled
until a vertical rod is achieved.  Instead we see at some point a
sharp first order transition between the highly stretched vertical rod
and a layered system with short tail.

In figures 2(b) and 2(c) we show plots of the logarithm of the maximum
eigenvalue of the matrix of second derivatives in the variables
$\beta_b$ and $\beta_s$ of $\log(Z_n(\beta_b,\beta_s,\beta_f))$ at
fixed $\beta_f$ (as in Figure 2(a)) but at values of $\beta_f$ being
$1.5$ and $3.0$. It is clear that as $\beta_f$ is increased the
stretched phase that occurs for small $\beta_b$ and $\beta_s$ expands
while the positions of the other phases and transitions move
little. We immediately note that these plots \emph{do not} tell the
whole story since physically one is usually interested in fixing the
force $f$ rather than $\beta_f$: fixing $\beta_f$ implies that the
force applied goes to zero at low temperatures. It is for this reason
that the re-entrant behaviour for absorbing polymers
\cite{orlandini2004a-a,krawczyk2004=a-:a} is not seen directly in
these plots. However, re-entrant behaviour does occur and occurs for
\emph{any} ratio of surface to bulk energies. Let us now consider the
more traditional force-temperature diagram and return to this point.

In figure 4 a plot of the force $f_c(T, \alpha)$ needed to pull a
polymer from the wall for a ranges of temperatures and a parameter
$\alpha$ which measures the relative strength of the surface (wall)
interaction to the self-interaction. 
\begin{figure}[ht!]
\begin{center}
\includegraphics[width=86mm]{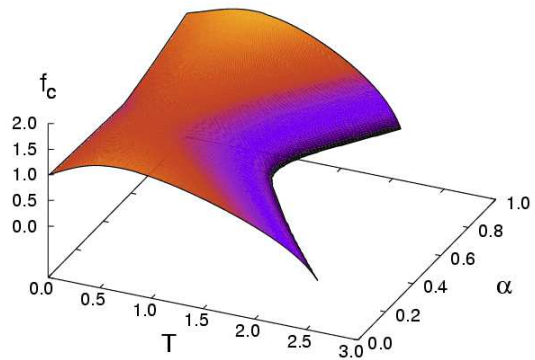}
\vspace{1mm}
\caption{\it A plot of the force $f_c$ needed to pull a polymer
from the surface against temperature $T$ and a parameter
$\alpha$. The parameter $\alpha$ controls the relative strength of
wall attraction and self-attraction with $\varepsilon_s=\alpha$ and
$\varepsilon_b = 1 - \alpha$. The limiting cases of surface
desorption and of pulling a collapsed polymer are easily visible in the
plot for $\alpha=1$ and $\alpha=0$, respectively.}
\end{center}
\end{figure}
We have parameterised the energies of surface
and self-attraction as $\varepsilon_s=\alpha$ and
$\varepsilon_b=1 - \alpha$ respectively. Using this parameterisation for
$0\leq \alpha \leq 1$ gives the whole range of attractive activities:
the ratio of surface to bulk activities is given as $\beta_s/\beta_b=
\alpha/(1 - \alpha)$ and so is constant for fixed $\alpha$. For $\alpha=0$
we have $\varepsilon_s=0$ and $\varepsilon_b=1$ which corresponds to
pure self-attraction while the other boundary of the parameter space
with $\alpha=1$ gives $\varepsilon_s=1$ and $\varepsilon_b=0$ which is
the pure surface adsorption case. This extends the diagrams given in
\cite{orlandini2004a-a,krawczyk2004=a-:a} in which only adsorption is
considered. If a force smaller than $f_c$ is applied the polymer is in
the phase appropriate to the value of $\alpha$: either collapsed or
adsorbed or both. On the other hand for forces larger than $f_c$ the
polymer is in the `stretched' phase. We immediately note that the
reentrant behaviour observed in the adsorption-only case
\cite{orlandini2004a-a,krawczyk2004=a-:a} persists for all $\alpha$.
Fixing the force to be at a value slightly larger than the zero
temperature critical force and then increasing the temperature leads
to transitions from the stretched state to a non-stretched phase and
back again to the stretched state. This arises due to the entropy of
the zero temperature state; one can easily extend the arguments in
\cite{orlandini2004a-a} to demonstrate that re-entrant behaviour can
occur even when the zero-temperature configuration of the
non-stretched state is a Hamiltonian (fully compact) walk in a cube
rather than a totally adsorbed polymer.

%Consider the phase diagram in the
%$\beta_b$-$\beta_s$ plane for fixed $f$.  There are ranges of $f$
%where the stretched phase appears at large $\beta_b$ and $\beta_s$
%moving around the boundary of the plane from the $\beta_b=0$ corner to
%the $\beta_s=0$ corner as $f$ is increased. In particular, consider a
%ray in the $\beta_b$-$\beta_s$ plane such that for $f=0$ and low
%temperatures the polymer is in a 2-layer phase for large but finite
%$N$. Then, there is some range of values of $f$ for which one would
%see a stretched phase, a layered phase and then another stretched
%phase as the temperature is raised from zero.

Let us discuss the re-entrant behaviour returning to the
($\beta_b$,$\beta_s$,$\beta_f$)-parametrisation. For fixed
$\varepsilon_b$, $\varepsilon_s$, and $f$, changing the temperature
$T$ implies moving on a ray in the
($\beta_b$,$\beta_s$,$\beta_f$)-space. At high temperatures the system
is in the stretched phase near the origin. At low temperatures, the
state of the system depends on the choice of $\varepsilon_b$,
$\varepsilon_s$, and $f$.  For very large $f$ the system remains
stretched at all temperatures, whereas for very small $f$ and low
temperatures, the system is in a layered phase.  The re-entrant
behaviour manifests itself in the following way: there is a range of
$f$ for which the system, when moving along a ray in the
($\beta_b$,$\beta_s$,$\beta_f$)-space (ie.\ for fixed $\varepsilon_b$,
$\varepsilon_s$, and $f$), changes from a stretched state near the
origin to a layered one at intermediate temperatures and then changes
back to a stretched state as $\beta_b$, $\beta_s$, and $\beta_f$
become larger.

If the critical force is zero then the curve in the $T-\alpha$ plane
corresponds to the phase boundary of the DE phase with the apex of the
curve around $\alpha \approx 1/2$ being the location of the
multicritical point. On the other hand for $T=0$ there is a kink in
the function $f_c(\alpha)$ at exactly $\alpha=1/2$ which is a
consequence of the first order point coming from the transition from
SAG/layer phases from small $\alpha$ to the AC phase for larger
$\alpha$. There is the appearance of a kink joining the multicritical
point to the zero temperature transition which is presumably a
finite temperature effect of the transition to the AC phase.

In this paper we have studied how the phase diagram of a
self-attracting polymer that is also attracted and tethered to a flat
wall changes as a vertical force is applied to the un-tethered end of
the polymer. We have accomplished this using a flat histogram Monte
Carlo simulation that is capable of studying the whole range of
microscopic energies, temperature and polymer length up to a maximum
of $91$ monomers. We demonstrate that re-entrant behaviour occurs at
low temperature and for a range of forces for \emph{all} relative
strengths of self-attraction and surface attraction. For small forces
we have found that only the transition boundary of the ``stretched''
phase moves with increasing force while the rest of the phase diagram
is relatively unchanged. In contradiction to what may be expected we
have found that the novel layering meta-phases found for large but
finite polymer length are unaffected by small forces.

\section*{Acknowledgements} 

Financial support from the DFG is gratefully acknowledged by JK and
TP. Financial support from the Australian Research Council is
gratefully acknowledged by ALO and AR. ALO also thanks the Institut
f\"ur Theoretische Physik at the Technische Universit\"at Clausthal.

%\begin{thebibliography}{22}
%\expandafter\ifx\csname natexlab\endcsname\relax\def\natexlab#1{#1}\fi
%\expandafter\ifx\csname bibnamefont\endcsname\relax
%  \def\bibnamefont#1{#1}\fi
%\expandafter\ifx\csname bibfnamefont\endcsname\relax
%  \def\bibfnamefont#1{#1}\fi
%\expandafter\ifx\csname citenamefont\endcsname\relax
%  \def\citenamefont#1{#1}\fi
%\expandafter\ifx\csname url\endcsname\relax
%  \def\url#1{\texttt{#1}}\fi
%\expandafter\ifx\csname urlprefix\endcsname\relax\def\urlprefix{URL }\fi
%\providecommand{\bibinfo}[2]{#2}
%\providecommand{\eprint}[2][]{\url{#2}}
%
%
%

%\end{thebibliography}

%\bibliography{refs-aleks-general,refs-aleks-personal}

\begin{thebibliography}{10}

\bibitem{strick2001a-a}
T.~Strick, J.-F. Allemand, V.~Croquette, and D.~Bensimon,
\newblock Phys. Today {\bf 54}, 46 (2001).

\bibitem{marenduzzo2003a-a}
D.~Marenduzzo, A.~Maritan, A.~Rosa, and F.~Seno,
\newblock Phys. Rev. Lett. {\bf 90}, 088301 (2003).

\bibitem{rosa2003a-a}
A.~Rosa, D.~Marenduzzo, A.~Maritan, and F.~Seno,
\newblock Phys. Rev. E. {\bf 67}, 041802 (2003).

\bibitem{orlandini2004a-a}
E.~Orlandini, M.~Tesi, and S.~Whittington,
\newblock J. Phys. A: Math. Gen. {\bf 37}, 1535 (2004).

\bibitem{gennes1979a-a}
P.-G. de~Gennes,
\newblock {\em Scaling Concepts in Polymer Physics},
\newblock Cornell University Press, Ithaca, 1979.

\bibitem{halperin1991a-a}
A.~Halperin and E.~B. Zhulini,
\newblock Europhys. Lett {\bf 15}, 417 (1991).

\bibitem{lai1996a-a}
P.~M. Lam,
\newblock Phys. Rev. E {\bf 53}, 3819 (1996).

\bibitem{grassberger2002a-a}
P.~Grassberger and H.~Hsu,
\newblock Phys. Rev. E {\bf 65}, 031807 (2002).

\bibitem{krawczyk2004=a-:a}
J.~Krawczyk, A.~L. Owczarek, T.~Prellberg, and A.~Rechnitzer,
\newblock cond-mat/0407611.

\bibitem{mishra2004a-a}
P.~Mishra, S.~Kumar, and Y.~Singh,
\newblock cond-mat/0404191.

\bibitem{celestini2004=a-a}
F.~Celestini, T.~Frisch, and X.~Oyharcabal,
\newblock cond-mat/0406187.

\bibitem{krawczyk2004=b-:a}
J.~Krawczyk, A.~L. Owczarek, T.~Prellberg, and A.~Rechnitzer,
\newblock submitted to PRL.

\bibitem{prellberg2004a-a}
T.~Prellberg and J.~Krawczyk,
\newblock Phys. Rev. Lett. {\bf 92}, 120602 (2004).

\bibitem{veal1991a-a}
A.~R. Veal, J.~M. Yeomans, and G.~Jug,
\newblock J. Phys. A {\bf 24}, 827 (1991).

\bibitem{vrbova1996a-a}
T.~Vrbov{\'{a}} and S.~G. Whittington,
\newblock J. Phys. A {\bf 29} (1996).

\bibitem{rajesh2002a-a}
R.~Rajesh, D.~Dhar, D.~Giri, S.~Kumar, and Y.~Singh,
\newblock Phys. Rev. E. {\bf 65}, 056124 (2002).

\bibitem{singh2001a-a}
Y.~Singh, D.~Giri, and S.~Kumar,
\newblock J. Phys. A. {\bf 34}, L67 (2001).

\end{thebibliography}
%\bibliographystyle{aip}

\end{document}